# Recovering Joy's Law as a Function of Solar Cycle, Hemisphere, and Longitude

B.H. McClintock[1] • A.A. Norton[1,2]

[1]Centre for Astronomy, James Cook University, Townsville, QLD, 4810 Australia

email: u1049686@umail.usq.edu.au

[2]HEPL, Stanford University, CA, 94305, USA

email: aanorton@stanford.edu

**Abstract** Bipolar active regions in both hemispheres tend to be tilted with respect to the East – West equator of the Sun in accordance with Joy's law that describes the average tilt angle as a function of latitude. Mt. Wilson observatory data from 1917 – 1985 are used to analyze the active-region tilt angle as a function of solar cycle, hemisphere, and longitude, in addition to the more common dependence on latitude. Our main results are as follows: i) We recommend a revision of Joy's law toward a weaker dependence on latitude (slope of 0.13 – 0.26) and without forcing the tilt to zero at the Equator. ii) We determine that the hemispheric mean tilt value of active regions varies with each solar cycle, although the noise from a stochastic process dominates and does not allow for a determination of the slope of Joy's law on an 11-year time-scale. iii) The hemispheric difference in mean tilt angles, 1.1° $\pm$ 0.27, over Cycles 16 to 21 was significant to a three-$\sigma$ level, with average tilt angles in the northern and southern hemispheres of 4.7° $\pm$ 0.26 and 3.6° $\pm$ 0.27 respectively. iv) Area-weighted mean tilt angles normalized by latitude for Cycles 15 to 21 anti-correlate with cycle strength for the southern hemisphere and whole-Sun data, confirming previous results by Dasi-Espuig, Solanki, Krivova, *et al.* (2010, *Astron. Astrophys.* 518, A7). The northern hemispheric mean tilt angles do not show a dependence on cycle strength. vi) Mean tilt angles do not show a dependence on longitude for any hemisphere or cycle. In addition, the standard deviation of the mean tilt is 29 – 31° for all cycles and hemispheres indicating that the scatter is due to the same consistent process even if the mean tilt angles vary.

## 1. Introduction

It is believed that magnetic fields generated at the base of the convective zone become buoyant and rise as toroidal flux tubes. Oriented in the East – West direction, flux-tube loops emerge from the solar surface to form sunspots. Observations of bipolar sunspots, on average, show leading spots closer to the Equator than following spots. Known as Joy's law, this was first published by Hale *et al.* (1919) after statistical analysis showed that the mean tilt angle of bipolar sunspots increased with latitude in both hemispheres. Joy's law has traditionally been interpreted as the Coriolis force operating in the separate hemispheres on motion in the rising magnetic-flux tubes. Coriolis forces dissipate once flux-tube emergence ends and tilt should relax to zero, but observations made by Howard (2000) showed tilt trending toward average, non-zero values after emergence. Babcock (1961) proposed that tilt is due to a spiral orientation of initial magnetic-field lines prior to emergence. Tilt-angle dependence on the latitude has been confirmed by many authors (Howard, 1991; Wang and Sheeley, 1991; Sivaraman, Gupta, and Howard, 1993, 1999). and provides constraints on the magnetic field strength of the flux tubes which emerge to form the observed active regions (D'Silva and Howard, 1993; Schüssler *et al.*, 1994).

We analyze the tilt angles independently by hemisphere. Since the transport of magnetic fields in Babcock - Leighton dynamo models (Babcock, 1961; Leighton, 1964, 1969) is partly achieved through a meridional-circulation cell seated in an individual hemisphere, the northern and southern hemispheres can become decoupled to some degree (Dikpati and Gilman, 2001; Chatterjee, Nandy, and Choudhuri, 2004). While it is obvious from the butterfly diagram that some degree of cross-hemispheric coupling prevents the hemispheres from becoming grossly out-of-phase (at least for solar cycles observed since the late 1800s), nevertheless, hemispheric phase lags are observed. For example, the polar-field reversals in the northern and southern hemispheres occurred half a year apart during Cycle 23 (Durrant and Wilson, 2002; Norton and Gallagher, 2010) and the northern hemisphere led the southern by 19 months in the declining phase of Cycle 20 (Norton and Gallagher, 2010) while hemispheres have been observed to be up to two years out of phase.

In addition to temporal phase lags between the hemispheres, it is common that one hemisphere dominates the other in the production of sunspot numbers and sunspot area (Temmer *et al.*, 2006). McIntosh *et al*. (2013) suggest that hemispheric asymmetry is a normal ingredient of the solar cycle and has important consequences in the structuring of the heliosphere. Charbonneau (2007) finds a "rich variety of behavior characterizing the two-hemisphere dynamo solution" including intermittency (a cessation of sunspot production similar to the Maunder Minimum) operating independently in separate hemispheres. Data analysis separated into hemispheres is critical to avoid blurring a signal that may be distinct in isolated hemispheres.

A tipping ($m = 1$ mode) or warping ($m > 1$ mode) of the toroidal magnetic band in the solar interior with respect to the equatorial plane in one or both hemispheres due to an MHD instability, as proposed by Cally, Dikpati, and Gilman (2003) and observed by Norton and Gilman (2005), would impart initial tilt angles dependent on longitude prior to a flux-rope's rise through the convection zone. An $m=0$ instability is expected for toroidal fields stronger than 50 kGauss on average whereas $m > 0$ is more likely for weaker toroidal fields. The growth of the tipping or deformation, and whether it is symmetric or asymmetric across the Equator, depends in part upon the width of toroidal band (Cally, Dikpati, and Gilman, 2003). A toroidal field tipped with respect to the Equator would not produce a different mean tilt angle averaged over longitude and latitude for a given cycle, but it would increase the scatter of the mean tilt angle. It could also explain why the tilt does not relax to zero after the active region has fully emerged as observed by Kosovichev and Stenflo (2008) and summarized nicely as follows: "It may be that Joy's law reflects not the dynamics of the rising flux tube, but the spiral orientation of the toroidal magnetic field lines below the surface as suggested by Babcock (1961)". We argue that Joy's law is due to a combination of both the Coriolis force's acting on the rising flux as it rises as well as an initial tilt imparted to the flux rope from the toroidal geometry that it retains. We search for a dependence of tilt angle on longitude as well as a dependence of noise in the mean tilt angle as a function of solar-cycle strength. It also appears that the tilt angle is inherently noisy, presumably due to the turbulent convection that is encountered by the flux ropes during their rise. However, Stenflo and Kosovichev (2012) argue that the many examples of large bipolar active regions with tilts that differ from the expected Joy's law angle by 90° are not simply regions buffetted by turbulent convection, but instead are regions from a different flux system that coexists at any given latitude.

## 2. Recovering Joy's Law

Furthering work by Howard (1996), and others, we examine bipolar active-region tilt angles observed at the Mt. Wilson observatory. We also record tilt angle dependence on hemisphere, solar cycle, latitude, and longitude (dependence on longitude discussed in Section 4). In Figure 1, mean tilt-angle values as a function of latitude for each hemisphere are shown averaged over Solar Cycles 16 to 21 for data collected at the Mt. Wilson observatory between 1923 and 1985. Cycle 15 began in 1913, but Mt. Wilson observations for this data did not begin until 1917, near solar maximum of this cycle. We excluded Cycle 15 from this analysis as it is an incomplete representation of a solar cycle. Mt. Wilson data are not available for the end of solar cycle 21 from January 86 to September 86. However, the monthly smoothed sunspot number had dropped to around 12.2 by January 86. At most, this would have amounted to approximately 110 spots *versus* the 4000 pairs in this cycle. After removing single sunspots from analysis, the effect on overall results would have been negligible. Cycle 21 is therefore included. The only regions excluded were individual spots, *i.e.* groups that did not have at least one sunspot in both the leading and following portions of a group. These were indexed in the Mt. Wilson data with a tilt angle of zero. Dasi-Espuig *et al.* (2010) thoroughly investigated entries with a zero tilt angle and found only one data point that corresponded to a true tilt value of zero whereas all others were single sunspots whose tilt angle could not be defined. The sample standard deviation of each latitudinal bin is divided by the square root of the bin population number and overplotted as standard error bars.

Empirical Joy's law equations from previous works are also plotted in Figure 1 as described by Wang and Sheeley (1991) as Equation (1), Leighton (1969) as Equation (2), Norton and Gilman (2005) as Equation (3), and Dasi-Espuig, *et al.* (2010) as Equation (4)

$$\sin \gamma = 0.48 \sin \theta + 0.03 \qquad (1)$$

$$\sin \gamma = 0.5 \sin \theta \qquad (2)$$

$$\gamma = 0.2\theta + 2.0 \qquad (3)$$

$$\gamma = (0.26 \pm 0.05)\theta \qquad (4)$$

where $\gamma$ is tilt angle and $\theta$ is latitude. Southern hemisphere latitudes are considered positive for plotting purposes. Tilt angles in both hemispheres are considered positive if the leading spot is closer to the Equator than the following spot.

In order to understand Equations (1) – (4), some background on data and analysis is in order. Equation (1) was formulated by Wang and Sheeley (1991) after analysis of National Solar Observatory/Kitt Peak data, utilizing 2710 magnetograms of bipolar magnetic regions (BMRs) collected during Solar Cycle 21. Tilt angles were determined by hand, analyzing magnetogram prints at a time of approximate peak flux for each BMR. Averages were flux-weighted and taken over sine-latitude bins of width 0.05 (approximately 3°). Equation (2) was formulated by Leighton (1969) who approximated Joy's law from measurements by Brunner (1930). Norton and Gilman (2005) implemented Joy's law as part of a sunspot-behavior model and Equation (3) is the best fit to an average of tilt angle as a function of latitude for over 650 active regions observed in Michelson Doppler Imager (MDI) data from 1996 – 2004 (Norton and Gilman, 2004). Equation (4) was determined by Dasi-Espuig *et al.* (2010) using available Mt. Wilson data, including the latter part of Solar Cycle 15 to most of Cycle 21. The data were binned by 5° latitude, area-weighted in an effort to reduce scatter, and linear fits were forced through the origin for Equation (4).

We find a linear fit for the relationship of the northern and southern average tilt angles as a function of latitude to be:

$$\gamma_N = 0.26\theta + 0.58 \text{ (or } \sin\gamma_N = 0.271\sin\theta + 0.010 \text{)} \quad (5)$$

$$\gamma_S = 0.13\theta + 1.38 \text{ (or } \sin\gamma_S = 0.425\sin\theta + 0.024 \text{)} \quad (6)$$

The values of the binned, average tilt angles observed at the higher latitudes are not well-fit by the Wang and Sheeley (1991) or previous historical Joy's law equations. We propose an updated Joy's law with a lower slope between 0.13 – 0.26, as seen in Equations (5) and (6) for the northern and southern hemisphere determined using Mt. Wilson data from Cycles 16 to 21. The linear correlation coefficient for Equations (5) and (6) are 0.96 and 0.65 for the northern and southern hemispheres respectively. The correlation coefficient (-1 $\leq r \leq$ 1) measures the strength of the linear relationship between two variables and is defined as the covariance of the two variables divided by the product of their standard deviations. Values of -1 and 1 indicate a perfect inverse or direct relationship, respectively. We correlate the binned, mean tilt angle and latitude.

The hemispheric, linear fits to Joy's law are more consistent with Dasi-Espuig *et al.* (2010) who report a lower slope value of 0.26 – 0.28 than with the equations from the 1990s and prior that had higher slopes. We also propose that the Joy's law equation should not be forced through the origin. It is reasonable that the slope reported by Dasi-Espuig *et al.* (2010) is higher than the slope reported here because they force the fit through the origin, which we do not. We justify our approach as being purely observational. If we did force the fit through the origin, our slopes in Equations (5) – (6) would increase to be 0.29 and 0.20 in the northern and southern hemispheres, respectively. The data here consistently demonstrate that Joy's law does vary by hemisphere. It is possible that the mechanisms responsible for tilt angles in each hemisphere have a canceling effect on tilt near the Equator and is therefore not an accurate indication of Joy's law by hemisphere.

The results of Dasi-Espuig *et al.* (2010) showing that tilt angle is variable as a function of solar cycle are noteworthy. Our initial attempts to recover Joy's law for each hemisphere and solar cycle were frustrating due to the fact that Joy's law only appears weakly (Figure 2). Cycle 16 had low population (< 25) in the first (0° – 3°) and last (27° – 30°) bins, resulting in large error bars. These bins were subsequently removed from all individual cycle plots for consistency. The data are poorly fit by a linear function in most cases. The linear correlation coefficients range from r = 0.18 (Cycle 17 North, Cycle 19 South) to *r* = 0.86 (Cycle 20 North). The large amount of scatter and high noise apparent in Joy's law is interesting because it indicates a stochastic process is competing with the mechanism that determines the tilt angles. The stochastic process dominating Joy's law on the short time-scale is considered to be turbulent convection imparting random tilt angles to the rising flux tubes (Weber, Fan, and Miesch, 2012; Fisher, Fan, and Howard, 1995). We agree with Dasi-Espuig *et al.* (2010) who state that "no clear difference could be determined between the slopes of Joy's law from cycle to cycle", as can be seen in Figure 2, and therefore we use the mean tilt value from each hemisphere for each cycle to analyze the hemispheric differences.

It is possible that the recovery of a mean bipolar region tilt angle and scatter for a given solar cycle can be used as a diagnostic for that cycle, *i.e.* the strength of the cycle as indicated by Dasi-Espuig *et al.* (2010) or the geometry/orientation of the toroidal fields from which the flux ropes begin their initial rise (Babcock, 1961, Norton and Gilman, 2004). Simulations by Weber, Fan, and Miesch (2012) of thin flux tubes rising through solar-like turbulent convections show how much tilt-angle scatter increases with decreasing flux and field strength. Therefore, quantifying the scatter in Joy's law can constrain the flux and field strength within the context of their model, *i.e.* a larger scatter is indicative of flux tubes dominated by convection instead of magnetic buoyancy. In addition, since smaller average tilt angles minimize the amount of active-region flux that becomes the poloidal field,

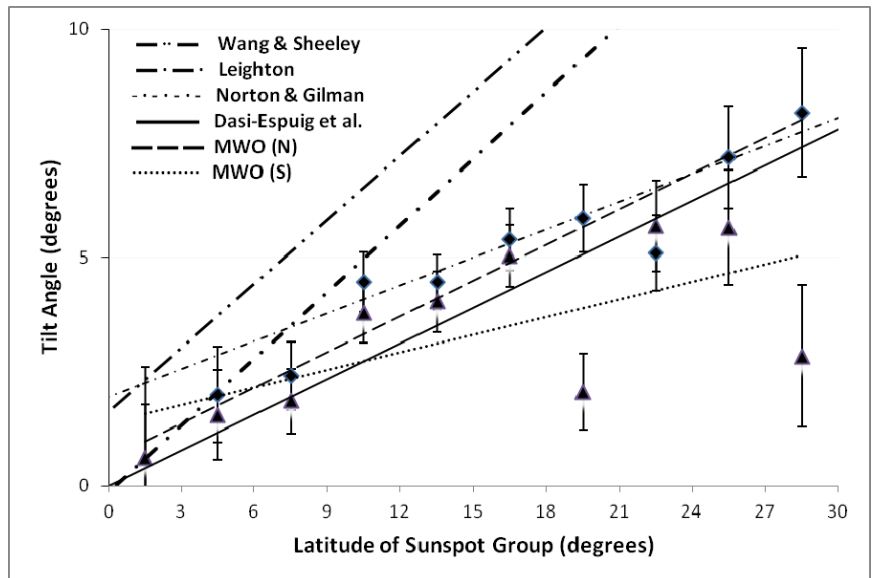

**Figure 1** Tilt angle as a function of latitude, northern (diamond) and southern (triangle) hemispheres, for Solar Cycles 16 to 21. Data were binned in 3° latitude. Standard error of the mean overplotted as error bars. Common Joy's law equations are plotted for reference Wang and Sheeley (1991), Equation (1), dashed-dot-dot; Leighton (1969), Equation (2), large dash-dot; Norton and Gilman (2005), Equation (3), small dash-dot; Dasi-Espuig, *et al.*, (2010), Equation (4), solid. Linear fit Equation (5) for northern hemisphere data (dash) and Equation (6) for southern hemisphere data (dot) also shown.

a smaller average tilt angle leads to a weaker polar cap mean field strength (Petrie, 2012).

We are uncertain why specific bins in the southern hemisphere showed such different behavior from the other bins. We found that late in all solar cycles (except 20) aberrant activity occurred at the 18° – 21° latitudes. In particular, the southern hemisphere during Cycle 19 is very disorganized, with the high-latitude bins of 18° – 21° (304 regions, 24 %) and 24° – 27° (150 regions, 12 %) having negative mean tilt values, meaning these bi-polar regions have a following spot closer to the Equator than the leading spot. The southern hemispheric tilt angles for Cycle 19 are responsible for the low mean tilt angles for the whole Sun in Cycle 19 as reported by Dasi Espuig *et al.,* (2010, see their Table 1). It would be of interest to study this in more detail and better understand the conditions favorable for aberrant configurations, *i.e.* anti-Hale and negative tilt angles, to occur.

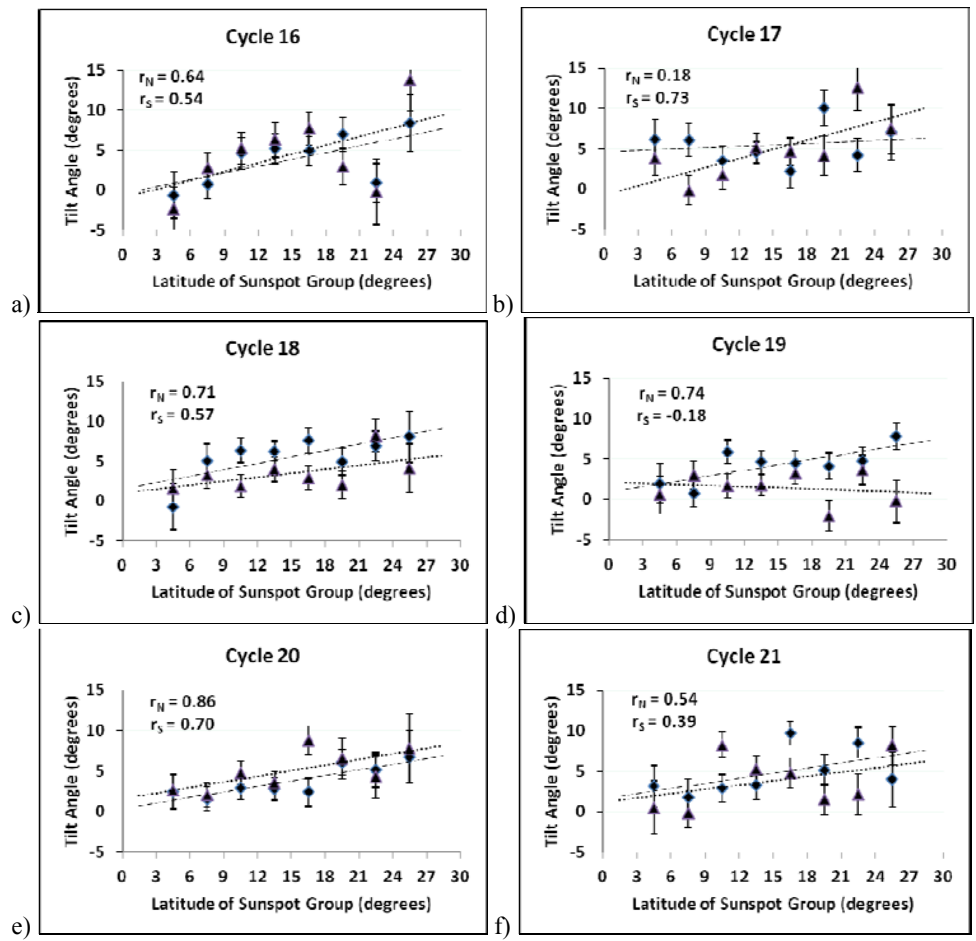

**Figure 2** Tilt angle as a function of latitude for the northern (diamond) and southern (triangle) hemispheres for Solar Cycles 16 to 21 are shown in panels (a) to (f), respectively. Data were binned in 3° latitude. Standard error of the mean overplotted as error bars. Linear fits to northern (dash) and southern (dot) hemisphere data are shown with linear correlation coefficients [$r_N$, $r_S$] included in the legends.

## 3. Joy's Law as a Function of Hemisphere

The average tilt angle and standard error of the mean for each hemisphere for Solar Cycles 16 to 21 are given in Table 1. Standard error of the mean was calculated as sample standard deviation divided by the square root of the sample number. Differences in Joy's law between hemispheres are poorly determined (below the two-σ level) for Cycles 16, 17, 20, and 21. However, Cycles 18 and 19 as well as the data averaged over all cycles show a significant difference between hemispheres. These findings are indicative that Joy's law varies by hemisphere and by solar cycle.

**Table 1** Mean Tilt Angle in degrees for northern hemisphere [$\bar{\gamma}_N$] and southern hemisphere [$\bar{\gamma}_S$] with standard deviation of mean [$\sigma_{\bar{\gamma}}$] for Solar Cycles 16 to 21. In addition, difference of hemispheric mean tilt angle and statistical significance are shown in last two columns.

| Solar Cycle | $(\bar{\gamma} \pm \sigma_{\bar{\gamma}})_N$ | $(\bar{\gamma} \pm \sigma_{\bar{\gamma}})_S$ | $\Delta\bar{\gamma} = |\bar{\gamma}_N - \bar{\gamma}_S|$ | $\dfrac{\Delta\bar{\gamma}}{\sqrt{\sigma^2_{\bar{\gamma}_N} + \sigma^2_{\bar{\gamma}_S}}}$ |
|---|---|---|---|---|
| 16 | 3.8° ± 0.73 | 4.4° ± 0.81 | 0.6 | 0.3 |
| 17 | 5.4° ± 0.70 | 4.0° ± 0.71 | 1.4 | 1.4 |
| 18 | 5.7° ± 0.61 | 2.9° ± 0.60 | 2.8 | 3.3 |
| 19 | 4.6° ± 0.53 | 1.8° ± 0.59 | 2.8 | 3.5 |
| 20 | 3.5° ± 0.60 | 4.8° ± 0.66 | 1.3 | 1.5 |
| 21 | 5.0° ± 0.67 | 4.4° ± 0.68 | 0.6 | 0.6 |
| 16 – 21 | 4.7° ± 0.26 | 3.6° ± 0.27 | 1.1 | 3.0 |

Using results in Table 1, we attempt to answer the following questions: Is there a significant difference between northern and southern hemispheric mean tilt? The last row of Table 1 indicates that yes, there is a significant difference of mean tilt at a three-σ level. Do the hemispheric differences in mean tilt values change from cycle to cycle? We find an average value of $\Delta\bar{\gamma}$ over all six cycles equal to 1.5 with a statistical significance of nearly four-σ (3.9). Therefore, we are convinced that there is significant variation in hemispheric mean tilts from cycle to cycle.

We agree with Dasi-Espuig *et al.* (2010) that a revision of Joy's law is necessary. Their conclusion that a relationship exists between cycle strength and mean tilt is intriguing and we attempted to confirm this result. We used the values reported by Goel and Choudhuri (2009) of total sunspot area in micro-hemispheres by solar cycle and hemisphere [$A_N$, $A_S$] for Cycles 15 through 21. Cycle 15 data were only available from just prior to solar maximum until the end of the cycle. The minimal effects of data missing from the last nine months of Cycle 21 are discussed in Section 2. Sunspot area is used as a proxy for cycle strength (Solanki and Schmidt, 1993). Areas were calculated from Royal Greenwich Observatory data. We compare total sunspot area to mean tilt separated by hemisphere and solar cycle in Table 2. Assuming larger total sunspot area indicates a stronger cycle and hemispheric differences exist within each cycle, we find evidence of the same inverse relationship as Dasi-Espuig *et al.* (2010) such that a stronger cycle produces less average tilt.

**Table 2** Sunspot Area [$10^4$ µ-hemispheres], mean tilt angle[$\bar{\gamma}$], mean tilt angle normalized by mean latitude $\bar{\gamma}/\bar{\lambda}$, and area-weighted mean tilt angle normalized by mean latitude $(\bar{\gamma}/\bar{\lambda})_\omega$ values are provided for northern and southern hemispheres and total Sun for Solar Cycles 15 – 21. The strength of the correlation of mean tilt with sunspot area was measured as the correlation coefficient [$r$] for each hemisphere and total-Sun values. Cycle 15 data were only available after solar maximum.

|  | Solar Cycle | | | | | | | |
| --- | --- | --- | --- | --- | --- | --- | --- | --- |
|  | 15* | 16 | 17 | 18 | 19 | 20 | 21 | $r$ |
| $A_N$ | 4.3 | 4.7 | 6.0 | 7.4 | 10.6 | 6.9 | 7.5 | |
| $\bar{\gamma}_N$ | 4.1° | 3.8° | 5.4° | 5.7° | 4.6° | 3.5° | 5.0° | 0.25 |
| $\bar{\gamma}_N/\bar{\lambda}$ | 0.35 | 0.27 | 0.36 | 0.37 | 0.26 | 0.24 | 0.33 | -0.17 |
| $(\bar{\gamma}_N/\bar{\lambda})_\omega$ | 0.45 | 0.29 | 0.50 | 0.46 | 0.30 | 0.32 | 0.40 | -0.29 |
| $A_S$ | 3.6 | 3.9 | 6.0 | 7.0 | 7.4 | 4.9 | 7.8 | |
| $\bar{\gamma}_S$ | 3.7° | 4.4° | 4.0° | 2.9° | 1.8° | 4.8° | 4.4° | -0.45 |
| $\bar{\gamma}_S/\bar{\lambda}$ | 0.27 | 0.32 | 0.28 | 0.20 | 0.12 | 0.35 | 0.29 | -0.67 |
| $(\bar{\gamma}_S/\bar{\lambda})_\omega$ | 0.43 | 0.43 | 0.27 | 0.29 | 0.11 | 0.33 | 0.28 | -0.83 |
| $A_{tot}$ | 7.9 | 8.6 | 12.0 | 14.5 | 18.0 | 11.9 | 15.3 | |
| $\bar{\gamma}_{tot}$ | 3.9° | 4.2° | 4.7° | 4.3° | 3.4° | 4.1° | 4.7° | -0.16 |
| $\bar{\gamma}_{tot}/\bar{\lambda}$ | 0.31 | 0.30 | 0.32 | 0.29 | 0.20 | 0.29 | 0.31 | -0.64 |
| $(\bar{\gamma}_{tot}/\bar{\lambda})_\omega$ | 0.44 | 0.35 | 0.39 | 0.38 | 0.23 | 0.32 | 0.34 | -0.75 |

In Figure 3, the area-weighted mean tilt values normalized by mean latitude (see $(\bar{\gamma}/\bar{\lambda})_\omega$ in Table 2) are plotted as a function of total sunspot area for Solar Cycles 15 to 21 for the northern and southern hemisphere as well as the total Sun. Dasi-Espuig *et al.* (2010) used area-weighting to give larger, and therefore, less scattered groups more influence on mean tilt. The mean latitude of sunspot emergence decreases and approaches zero as the solar cycle progresses. Normalizing by latitude removes that latitudinal bias and allows for the inclusion of incomplete cycles in our analysis. Linear regression lines are fit to normalized mean tilt and sunspot area for each hemisphere. Correlation coefficients [$r$] are found to be $r_N = -0.29$, $r_S = -0.83$, $r_{tot} = -0.75$ for the northern hemisphere, southern hemisphere, and total-Sun values.

There is an inverse correlation of area-weighted mean tilt to sunspot area and, by proxy, cycle strength in the southern hemisphere. The probability is 2.1 % that the linear correlation coefficient of $r_S = -0.83$ in the south is due to chance. Total-Sun values also suggest an inverse relationship of area-weighted mean tilt angle values with cycle strength, the correlation coefficient $r_{tot}$ = -0.75 having a 5.0% probability of chance. The correlation between mean tilt and cycle strength in the Northern hemisphere is insignificant. The smallest chance probabilities of 2.1 % and 5.0 % for the Southern hemisphere and total-Sun correlations are at or below the usual significance level of 5 %, and therefore we confirm a statistically significant negative correlation between area-weighted mean tilt value and cycle strength as measured by sunspot area in the southern hemisphere and the whole-Sun data.

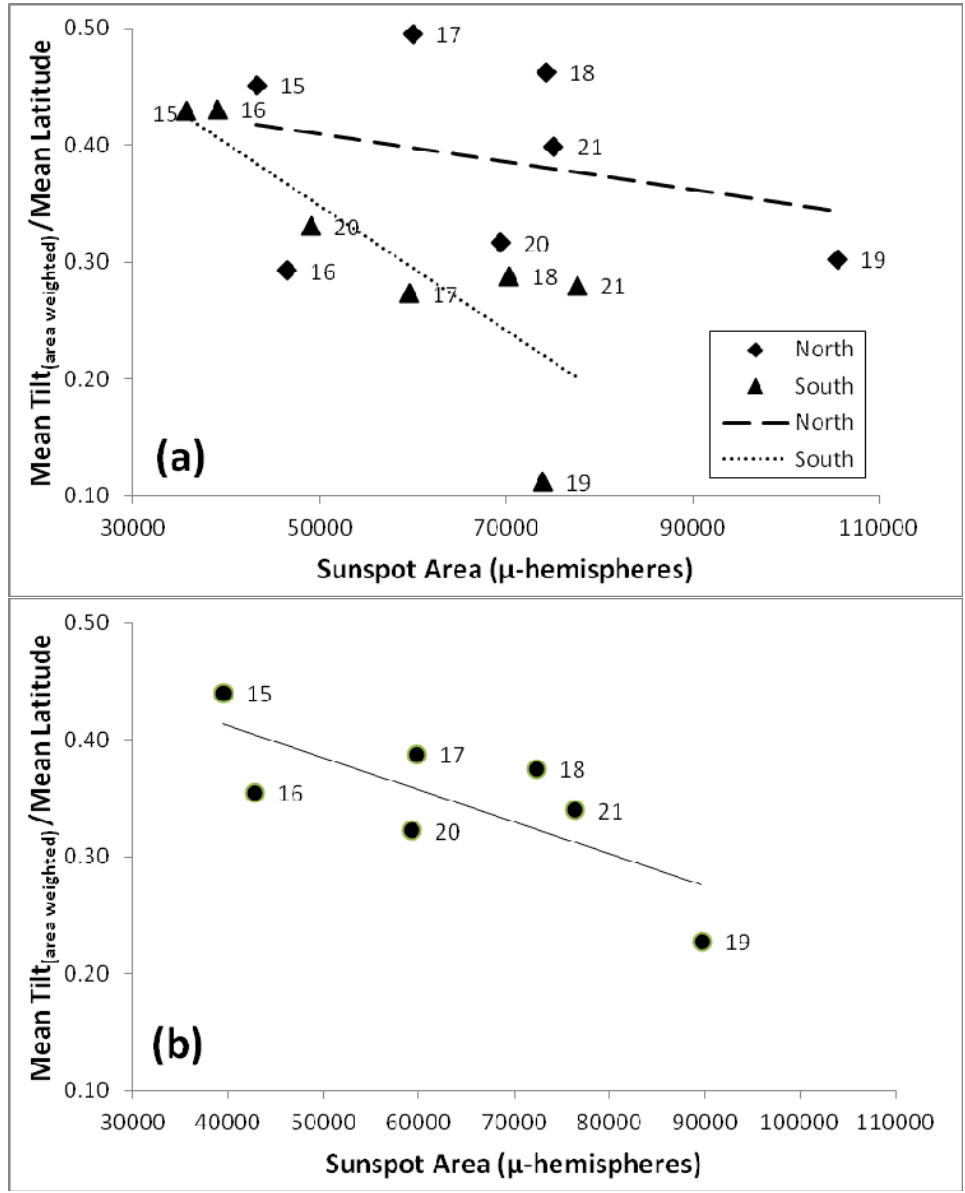

**Figure 3** Area-weighted mean tilt angle normalized by mean latitude (degrees) as a function of sunspot area [µ-hemispheres] for Cycles 15 to 21. Panel (a) shows northern hemisphere (diamond, dashed line) and southern hemisphere (triangle, dotted line), and panel (b) shows the total-Sun with sunspot area divided by two. Linear correlation coefficients, r, for each hemisphere and total-Sun are $r_N$ = -0.29, $r_S$ = -0.83, $r_{tot}$ = -0.75

## 4. Joy's Law as a Function of Longitude: Searching for Evidence of a Tipped Toroidal Field in Tilt-Angle Data

If toroidal magnetic fields at the base of the convection zone in each hemisphere were tipped with respect to the equatorial plane as proposed in theory by Cally, Dikpati, and Gilman (2003) and observations (Norton and Gilman, 2005), then flux tubes would begin their rise through the convection zone with a tilt dependent on longitude. This might be observable as a pattern when tilt angles in each hemisphere are studied as a function of longitude. It is well-established that active longitudes appear during each solar cycle and certain longitudes host active regions repeatedly over time (De Toma, White, and Harvey, 2000). If an $m = 1$ instability were present we would expect to see a sinusoidal pattern.

To reveal longitudinal structure, possibly relating to the orientation of the toroidal field in each hemisphere, we separated tilt data by hemisphere and solar cycle. Active-region tilt angles as a function of longitude were plotted for northern and southern hemispheres for all solar cycles, with data binned into 20° longitudes, then averaged. Plots for Cycles 18 – 20 are presented in Figure 4 (a) – (f). We expected an $m = 1$ sinusoidal pattern suggestive of a tipped toroidal field in each hemisphere. We attempted to fit the data with sinusoidal curves representing $m = 1$ through $m = 8$ patterns with various amplitudes. No fit to the data was statistically significant. Therefore, we report no longitudinal dependence in Joy's law.

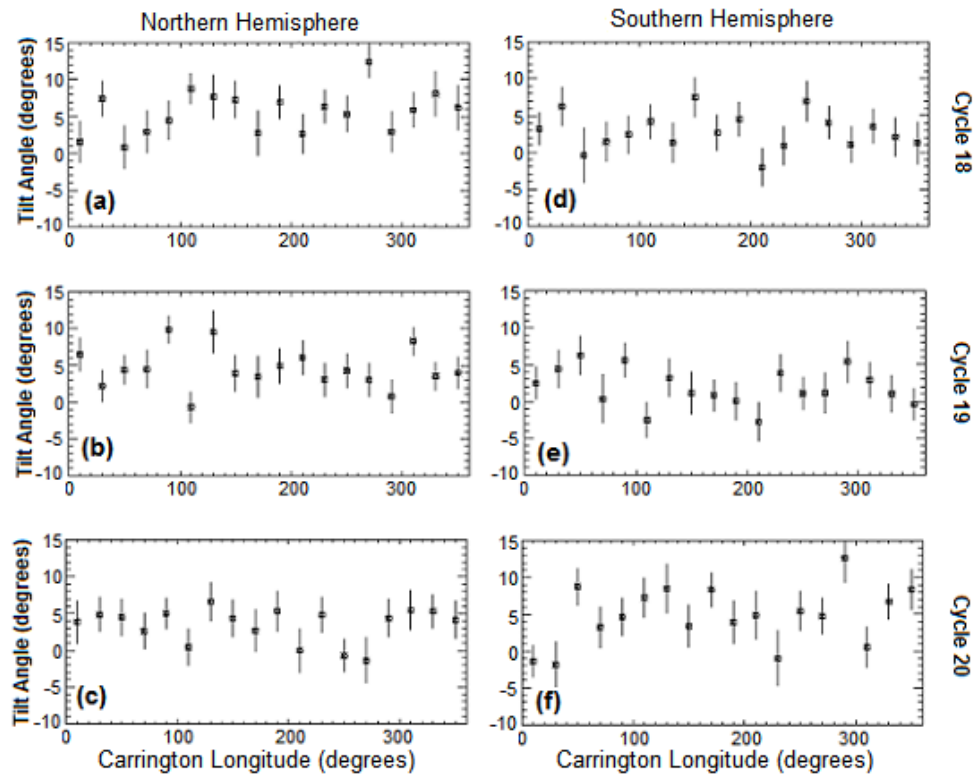

**Figure 4** Tilt angle as a function of longitude, Solar Cycles 18 – 20, northern (a) – (c) and southern hemispheres (d) – (f). Data were binned 20° in longitude. Standard deviations of tilt angle for all sunspot groups in each longitudinal bin are overplotted as error bars.

If a tipped toroidal field were only present for one to two years during a solar cycle, this might prevent a tilt angle dependence on longitude to be decipherable when averaging over ≈11 years. However, it may be possible to see increased scatter in the tilt-angle values for a cycle that has a tipped toroidal field compared to a cycle without one. For this reason, we determined the standard deviation (not the standard deviation of the mean) for the average tilt angle as a function of hemisphere and cycle (see Table 3). The standard deviation values have a very small range, from 29.3 – 31.2°, even though the strength of the cycle, shown as sunspot area, varies a great deal. The errors of the standard deviation values shown in Table 3 range from 0.53 – 0.81°. The small range of standard deviation values indicates that the source of scatter in the tilt angles is due to a process that is nearly identical from one cycle and hemisphere to the next. The values shown in Table 3 do not support the presence of a tilting or deformation of the toroidal band in the solar interior since there is no difference in scatter of observed tilt angles at the surface between one cycle and the next. We agree with Fisher, Fan, and Howard (1995) that the very small range of the standard deviations of the tilt angle (referred to as rms tilt in their paper) are consistent with a process such as the buffeting by convection which is persistent in scale as a function of longitude and latitude and similar from one cycle to the next.

**Table 3** Cycle strength in terms of sunspot area [$10^4$ µ-hemispheres], and the standard deviation, [$\sigma_\gamma$], (not the standard deviation of the mean) in the mean tilt angle are shown for the bipolar regions of the northern and southern hemisphere for Cycles 16 – 21.

| Cycle Strength and Standard Deviation of Average Tilt Angle | | | | | | |
|---|---|---|---|---|---|---|
| Cycle | 16 | 17 | 18 | 19 | 20 | 21 |
| $A_N$ | 4.7 | 6.0 | 7.4 | 10.6 | 6.9 | 7.5 |
| $\sigma_\gamma$ | 29.6° | 30.5° | 30.0° | 31.2° | 29.6° | 29.8° |
| $A_S$ | 3.9 | 6.0 | 7.0 | 7.4 | 4.9 | 7.8 |
| $\sigma_\gamma$ | 30.6° | 31.0° | 30.4° | 29.6° | 29.3° | 29.7° |

## 5. Summary

We determined that the mean tilt angle observed in Solar Cycles 16 to 21 was significantly different in the northern and southern hemispheres. Hemispheric differences up to 2.8° in average tilt angle persist across solar cycles. We suggest a revision to Joy's law equations with a weaker dependence on latitude (slopes of 0.26 and 0.13 for the northern and southern hemispheres were found) and more attention paid to the differences between hemispheres and cycles. We did not force the linear fit through the origin as Dasi-Espuig *et al.* (2010) did in their analysis. It is possible that bipolar active regions at the Equator have mean tilt angles of zero because the sampling is an aggregate of flux activity from both hemispheres. If we do force the fits through the origin, we find slopes of 0.29 and 0.20 for the northern and southern hemispheres, compared to 0.26 reported by Dasi-Espuig *et al.* (2010). Weber, Fan, and Miesch (2012) simulate rising flux tubes, including the effect of convection, and produce an expected slope for Joy's law dependent upon the strength of the source toroidal field and total flux in the tube. Our slope values of 0.29 and 0.20 in the northern and southern hemispheres are consistent with field strengths of 15 kG in the interior and flux ropes containing between $10^{20}$ and $10^{21}$ Mx.

We confirm the results of Dasi-Espuig *et al.* (2010) that whole-Sun mean tilt angles, weighted by area and normalized by latitude, for Cycles 16 to 21 show a statistically significant negative correlation with cycle strength (see Figure 3b). A tilt angle dependence upon cycle strength is a feedback mechanism in which the Sun can regulate sunspot-cycle amplitudes, *i.e.* a stronger cycle produces a smaller tilt angle and therefore a weaker poloidal seed field for the *n+1* cycle (Cameron and Schüssler, 2012). Jiang *et al.* (2010) study the effect of meridional flow perturbations and find that larger perturbations reduce the tilt angle of bipolar magnetic regions and thus diminish its contribution to the polar field. The perturbations are caused by near-surface inflows towards the active region band in each hemisphere, and the perturbation amplitude increases with stronger magnetic cycles. This mechanism may explain the observed anti-correlation between tilt angle and cycle strength. However, some doubts are cast on the results because the northern hemisphere did not exhibit a statistically significant negative correlation with cycle strength while the southern hemisphere did (see Figure 3a). We hope this result reinforces the importance of isolating data by hemisphere.

We searched for a non-axisymmetric mechanism at work by analyzing tilt angles as a function of longitude (see Figure 4 for Cycles 18 – 20). We attempted to fit the data with sinusoidal curves representing $m = 1$ through $m = 8$ patterns with various amplitudes. No fit to the data was statistically significant. Therefore, we find no evidence that tilt angles vary regularly in longitude. A toroidal field tipped with respect to the East – West direction would introduce a significant scatter into Joy's law if the flux rope retained some of the original tilt imparted to it from the source toroidal field. Therefore, we calculated the standard deviation of the average tilt angle from each cycle and hemisphere. The values exhibited a narrow range from 29.3 – 31.2° even though the cycle strengths varied greatly (see Table 3). This does not support the presence of a tilting or deformation of the toroidal field but is consistent with a process such as the buffeting by convection that is persistent in scale in latitude and longitude and similar from one cycle to the next.

Moreover, a bias toward reporting positive results regarding Joy's law may have impeded progress on this topic that would benefit from identifying time periods in which Joy's law cannot be recovered. These would be times in which the stochastic processes of turbulent convection dominate the tilt-producing mechanism thought to be the Coriolis force. The work by Weber, Fan, and Miesch (2012) is a great step towards the ability to interpret the scatter of bipolar region tilt angles in any period of the solar cycle to constrain the toroidal field strength in the interior and the flux residing in the thin flux tubes. The standard deviation values of the average tilt angle shown in Table 3 are consistent with Weber, Fan, and Miesch (2012) simulations of flux tubes containing $10^{21}$ Mx and forming from a toroidal field with the strength of 50 kG.